\documentclass[10pt,aps,prl,twocolumn,showpacs,superscriptaddress,nobalancelastpage]{revtex4-1}

\usepackage{graphicx}
\usepackage{amsmath,amssymb,amsfonts,dsfont,accents}
\usepackage{color}
\usepackage[normalem]{ulem} 
\usepackage{multirow}
\usepackage{array}
\usepackage{upgreek}
\usepackage{mathrsfs}
\usepackage{newcommands}

\newcommand{\prlsection}[1]{\textit{#1.---}}

\begin{document}

\title{Ramsey Interferometry in Correlated Quantum Noise Environments}

\author{F\'elix Beaudoin}
\affiliation{$\mbox{Department of Physics and Astronomy, Dartmouth College, 6127 Wilder Laboratory, 
Hanover, New Hampshire 03755, USA}$}
\author{Leigh M. Norris}
\affiliation{$\mbox{Department of Physics and Astronomy, Dartmouth College, 6127 Wilder Laboratory, 
Hanover, New Hampshire 03755, USA}$}
\author{Lorenza Viola}
\affiliation{$\mbox{Department of Physics and Astronomy, Dartmouth College, 6127 Wilder Laboratory, 
Hanover, New Hampshire 03755, USA}$}

\date{\today}

\begin{abstract}
We quantify the impact of spatio-temporally correlated Gaussian quantum noise on frequency estimation by Ramsey interferometry. While correlations in a classical noise environment can be exploited to reduce uncertainty relative to the uncorrelated case, we show that quantum noise environments with frequency asymmetric spectra generally introduce additional sources of uncertainty due to uncontrolled entanglement of the sensing system mediated by the bath. For the representative case of collective noise from bosonic sources, and experimentally relevant collective spin observables, we find that the uncertainty can increase exponentially with the number of probes. As a concrete application, we show that correlated quantum noise due to a lattice vibrational mode can preclude superclassical precision scaling in current amplitude sensing experiments with trapped ions.
\end{abstract}

\maketitle

A chief aim in quantum metrology is to demonstrate an advantage over classical approaches in the scaling of the precision to which a physical parameter may be estimated as a function of the number $N$ of probes being used (qubits in the simplest case)~\cite{metrology}. The use of entangled states yields asymptotic precision bounds which surpass the optimal $N^{-1/2}$ scaling achievable classically (the {\em standard quantum limit}, SQL), with the ultimate $N^{-1}$ precision bound set by the {\em Heisenberg limit}. Such superclassical scalings can benefit tasks as diverse as frequency estimation~\cite{bollinger1996optimal}, magnetometry~\cite{magnetometry}, thermometry~\cite{stace2010quantum}, force and amplitude sensing~\cite{biercuk2010ultrasensitive,gilmore2017amplitude}.  Prominent applications include gravitational-wave detection~\cite{grote2013first} and high-precision timekeeping with atomic clocks~\cite{clocks}, with a growing role being envisioned in biology~\cite{taylor2016quantum}.

Realizing the full potential of quantum metrology demands that the impact of realistic noise sources be quantitatively accounted for.  While no superclassical scaling is permitted under noise that is temporally uncorrelated and acts independently on each probe \cite{nocorrelation}, {\em noise correlations} can be beneficial in restoring metrological gain. For spatially uncorrelated noise, temporal correlations may be exploited to achieve a superclassical (Zeno-like) scaling at short detection times~\cite{chin2012quantum, matsuzaki-smirne}. For temporally uncorrelated noise, spatial correlations may enable superclassical scaling via a decoherence-free subspace encoding~\cite{dorner2012quantum,jeske2014quantum}, or they can be leveraged to filter noise from signal in quantum error-corrected sensing~\cite{layden2018spatial}. Even in the presence of simultaneous spatial and temporal correlations, as arising if the probes couple to a {\em common} environment with a {\em colored} spectrum, memory effects can be used to retain enhanced sensitivity over longer times, as long as the environment is modeled as {\em classical} \cite{szankowski2014parameter}.

The occurrence of non-trivial temporal correlations has been verified across a variety of systems through quantum noise spectroscopy experiments \cite{bylander2011noise,alvarez2011measuring,muhonen2014storing,malinowski2017spectrum,wang2017single,frey2017application,morello2018}; in typical metrological settings, spatial noise correlations also tend to naturally emerge due to probe proximity 
\cite{dorner2012quantum,monz2011fourteen}. Further, recent experiments have directly probed {\em non-classical} noise environments \cite{nonclassical}. The latter are distinguished by {\em non-commuting} degrees of freedom which translate, in the frequency domain, to spectra that are asymmetric with respect to zero frequency \cite{clerk2010introduction,paz2017multiqubit}. Crucially, qubits coupled to a {\em common, quantum} environment can become entangled in an uncontrolled way, leading to an additional source of uncertainty in parameter estimation that has not been accounted for to the best of our knowledge.  Such noise-induced entanglement is especially relevant to quantum metrology with spin-squeezed states generated by coupling qubits to common bosonic modes~\cite{bohnet2016quantum,vladan}, as this opens the door to correlated quantum noise due to vibrational~\cite{sawyer2012spectroscopy} or photonic sources~\cite{rigetti2012superconducting}. 

In this Letter, we provide a unified approach to Ramsey metrology protocols
under correlated quantum noise, by building on a transfer filter-function formalism \cite{paz-silva2014general} recently employed for control and spectral estimation of Gaussian quantum noise in multiqubit systems~\cite{paz2016dynamical,paz2017multiqubit}. We contrast the precision limits achievable with $N$ qubits initialized in a classical coherent spin state (CSS) and an experimentally accessible entangled one-axis twisted spin-squeezed state (OATS)~\cite{kitagawa1993squeezed,brask2015improved,bohnet2016quantum}. In the paradigmatic case of a collective spin-boson model, we find that the simultaneous presence of spatial and temporal correlations introduces a contribution to the uncertainty that grows exponentially with $N$, makes the precision scaling worse than SQL for a CSS, and prevents the SQL from being surpassed by use of a non-classical OATS. We further discuss a source of correlated quantum noise that has thus far been neglected in quantum-limited amplitude sensing with trapped ions~\cite{gilmore2017amplitude}. We argue that the resulting uncertainty can become dominant and preclude the realization of a superclassical scaling in this context.

\prlsection{Noisy Ramsey interferometry: Setting} We consider $N$ qubit probes, with associated Pauli matrices 
$\{ \s_n^{\alpha} \},$ $\alpha \in \{x,y,z\}$, $n=1, \ldots N$, each longitudinally coupled to a quantum bath through 
a bath operator $B_n$.
In the interaction picture with respect to the free bath Hamiltonian, $H_\mrm B$, we consider a joint Hamiltonian
of the form 
\begin{align}
H_\mrm {SB}(t)= \frac{\hbar}{2} \sum_{n=1}^N \, [{y_0(t)b+y(t)B_n(t)}] \, \s_n^{z},	
\label{eqnSB}
\end{align}
where $b$ is the angular frequency we wish to estimate, $B_n(t) \equiv \eul{i H_\mrm B t/\hbar}B_n\eul{-i H_\mrm B t/\hbar}$, 
and we allow for the possibility of open-loop control modulation via time-dependent functions $y_0(t), y(t)$. We assume that the initial joint state is factorized, $\rho_\mrm {SB} (0) \equiv \rho_0\otimes\rho_\mrm B$, and that the noise process described by $\{ B_n(t)\}$ is stationary and Gaussian with zero mean relative to $\rho_\mrm B$ \cite{paz2016dynamical}. Noise correlations 
are captured by the two-point correlation functions, $C_{nm}(t)\equiv \langle B_n(t) B_m(0) \rangle_\mrm B = \mrm{Tr_B}[ B_n(t) B_m(0) \rho_\mrm B]$, with the limiting cases of temporally or, respectively, spatially uncorrelated noise corresponding to $C_{nm}(t) = c_{nm} \delta(t)$ and $C_{nm}(t) = \delta_{nm} f_n(t)$. Coupling to a classical bath is recovered by letting $\{B_n(t)\}$ be commuting random variables,  
$[B_n(t), B_m(0)]_- \equiv 0$, $\forall m,n,t$.
In the frequency domain, the Fourier transform of $C_{nm}(t)$ yields the noise spectra, $S_{nm}(\omega)$. If 
$S_{nm}(\omega) \equiv \frac{1}{2}[ S_{nm}^+(\omega) + S_{nm}^-(\omega)]$, then 
${S^{\pm}_{nm}(\w) \equiv \int_{-\infty}^{\infty}dt\,\eul{-i\w t}\moy{ [B_n(t),B_m(0)]_{\pm}}}_\mrm 
B = S_{nm}(\omega) \pm S_{mn}(-\omega) $ define the ``classical'' ($+$) and ``quantum'' ($-$) spectra, respectively
~\cite{paz2017multiqubit}. By definition, quantum spectra vanish whenever noise is classical.  

\begin{figure*}
\begin{center}
\includegraphics[width=0.94\textwidth]{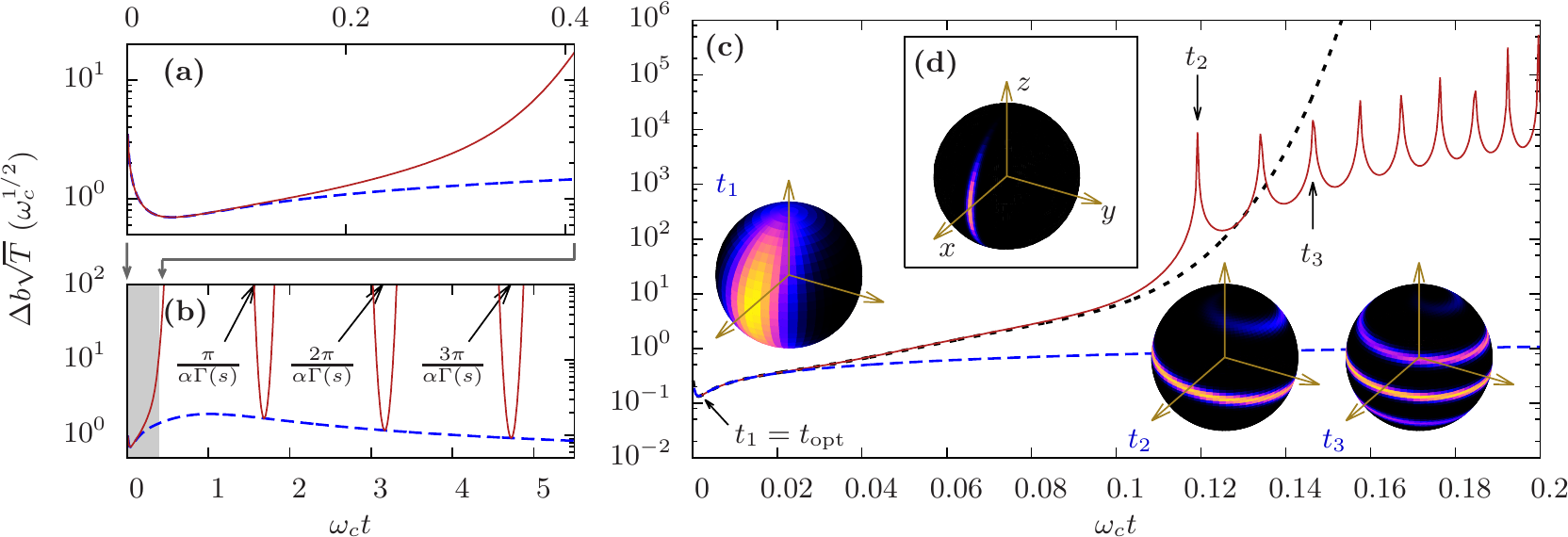}
\end{center}
\vspace{-3mm}
\caption{(Color online)
Detection-time dependence of the sensitivity $\D b (t)\sqrt T$ of parameter estimation in a collective spin-boson model, in units of $\omega_c^{\raisebox{-1pt}{\tiny{1/2}}}$, with $\omega_c$ being the upper angular frequency cutoff. (a) Initial CSS, short-time behavior. Solid (red) line: sensitivity with $\Y(t)\neq0$ resulting from the spin-boson model. Dashed (blue) line: sensitivity $\D b_0(t)\sqrt T$ for $\Y(t)=0$. Parameters: $\al=1$, $s=3$, $N=100$. (b) Initial CSS, long-time behavior. Parameters unchanged. Shaded area: short times. (c) Initial OATS. Solid (red) line: exact numerical calculation with $\Psi(t)\neq0$ from the spin-boson model. Dotted (black) line: cumulant expansion over the system (see text).
Dashed (blue) line: Exact calculation with $\Y(t)=0$. Insets: $Q$-functions for the system state at times $t_1$, $t_2$, and $t_3$ labeled in the main plot. Parameters: $\al=1$, $s=3$, $N=1000$. (d) $Q$-function corresponding to an initial OATS with minimal variance along $y$ for $N=1000$.\label{figspinboson}}
\end{figure*}

Starting from an arbitrary initial state $\rho_0$ that is not stationary under $H_\mrm {SB}(t)$, 
the resulting phase evolution can be detected through $\nu$ independent measurements of the collective spin 
$J_y\equiv \sum_n  \s_n^{y}/2$ (in units of $\hbar$). In particular: (i) $\rho_0 =\rho_{+\mvec {\hat x}}\equiv \proj+^{\otimes N}$ 
for an initial CSS, with $\ket\pm_n$ being $\pm 1$-eigenstates of $\s^x_{n}$; (ii) $\rho_0=\rho_\mrm{sq} \equiv U_\mrm{sq}\rho_{+\mvec{\hat x}}U_\mrm{sq}^\dag$ for an initial OATS, with $U_\mrm{sq} \equiv \eul{-i\bt J_x}\eul{-i\q J_z^2/2}$, and $\bt$ and $\q$
being rotation and twisting angles, respectively~\cite{kitagawa1993squeezed,ma2011quantum}. 
To quantify the precision in estimating $b$, we use the standard deviation~\cite{wineland1994squeezed}
\begin{align}
\!\!\!\D b (t) \equiv \frac{\nu^{-1/2}\D J_y (t)}{|\partial\moy{J_y(t)}/\partial b|}, \;\,
\D J_y^2 (t) \equiv \moy{J_y^2(t) }-\moy{J_y(t)}^2.	
\label{eqnuncertainty}
\end{align}
In a noiseless scenario ($B_n(t)\equiv 0$, $\forall n, t$), Ramsey interferometry yields an optimal uncertainty at the SQL, 
$\D b\propto N^{-1/2}$, with an initial CSS~\cite{wineland1994squeezed}, 
whereas an initial OATS with minimal uncertainty along $y$ [see Fig.~\ref{figspinboson}(d)] 
yields the superclassical scaling $\D b\propto N^{-5/6}$~\cite{kitagawa1993squeezed}.

\prlsection{Noisy Ramsey interferometry: Results} Since $H_\mrm{SB}(t)$ in Eq.~\eq{eqnSB} generates pure-dephasing dynamics, we may evaluate $\moy{\s_n^{y}(t)}$ and $\moy{\s_{n}^{y}\s_m^{y}(t)}$ by invoking the exact result in terms of generalized cumulants of bath operators established in Ref.~\cite{paz2017multiqubit}. Summing over all qubits and tracing out the bath, we then obtain, 
for arbitrary $\rho_0$~\cite{Supplement}, 
\begin{widetext}
\begin{align}
\!\!\!\!\moy{J_y(t)}&=\!\sum_{n} \eul{-\chi_{nn}(t)/2}\,\Tr_\mrm S\Big[\eul{-i\F_n(t)}\rho_0\frac{\s_n^{y}}2\Big],	\quad
\moy{J_y^2(t)}=\frac N4+\!\!\sum_{n,m\neq n}\!\!\!\eul{-[\chi_{nn}(t)+\chi_{mm}(t)]/2} \,\Tr_\mrm S
\Big[\eul{-i \F_{nm}(t)}\rho_0\frac{\s_n^{y}\s_m^{y}}4\Big], 
\label{eqnmomentsJy}\\
\!\!\F_n(t)&= \vf(t)\s_n^{z}+\!\!\sum_{\ell,\ell\neq n}\!\!\Y_{n\ell}(t)\s_n^{z}\s_{\ell}^{z},  \;\; 
\F_{nm}(t)\!= \vf(t) (\s_n^{z}+\s_m^{z})-i\ch_{nm}(t)\s_n^{z}\s_m^{z}+\!\!\!\!\sum_{\ell,\ell\neq nm}\!\!\left[\Y_{n \ell}(t)\s_n^{z}\s_\ell^{z}+
\Y_{m \ell}(t)\s_m^{z}\s_\ell^{z}\right]. 
\label{eqnevolution}
\end{align}
\end{widetext}
Above, we have introduced $\vf(t) \equiv b\int_0^t ds \, y_0(s)$, and effective propagators $\exp[-i\Phi_n(t)]$, $\exp[-i\Phi_{nm}(t)]$ 
that depend on two sets of real quantities: the decay parameters, $\chi_{nm}(t)$, describing loss of coherence, 
and the phase parameters, $\Y_{nm}(t)$, which characterize entanglement and squeezing mediated by the quantum bath. 
Explicitly,
\begin{align}
\ch_{nm}(t)&\equiv \frac1{2\pi}\mrm{Re}\int_{0}^{\infty}d\w\,F^+(\w,t)\,S_{nm}^+(\w),	
\label{eqnchi}\\
\Y_{nm}(t)& \equiv \frac1{2\pi}\mrm{Im}\int_{0}^{\infty}d\w\,F^-(\w,t)\,S_{nm}^-(\w),	
\label{eqnPsi}
\end{align}
where $F^+(\w,t) \equiv |\int_0^tds\,y(s)\eul{-i\w s}|^2$ and $F^-(\w,t) \equiv \int_0^tds\,y(s)\int_0^sdu\,y(u)\,\eul{-i\w(u-s)}$ 
are first- and second-order filter functions describing the action of $y(t)$ in the frequency domain. 
Clearly, $\Y_{nm}(t)\equiv 0$ if noise is classical. 

For illustration, we assume henceforth a {\em collective} noise regime, $B_n(t)\equiv B(t)$ $\forall n,t$, by 
deferring a more complete analysis to a separate investigation \cite{Next}. Thus, $\chi_{nm}(t)\equiv \chi(t)$, $\Psi_{nm}(t)\equiv \Psi(t)$. A non-zero phase parameter $\Y(t)\neq0$ is then distinctive of quantum noise that is both spatially and temporally correlated~\cite{Classical}. 

(i) {\em Initial CSS}. Since such an initial state is separable, we can evaluate $\mean{J_y(t)}$ and $\mean{J_y^2(t)}$ exactly. Substituting into Eq.~\eq{eqnuncertainty}, and minimizing the resulting uncertainty with respect to $b$ by taking $\vf=k\pi$, $k\in\mathbb N$, we find~\cite{Supplement}
\begin{align}
\!\!\!\!   \D b(t)^2 \!=\!\frac{(N+1)\eul{\chi(t)}-(N-1)\eul{-\chi(t)}\cos^{N-2}2\Y(t)}{2N\nu [\int_0^tds\,y_0(s)]^2\cos^{2N-2}\Y(t)}.	
\label{eqnCSS}
\end{align}
Note that $\D b$ is periodic with respect to $\Y$, in the sense that $\D b(\Y+\pi)$=$\D b(\Y)$. In addition, Eq.~\eq{eqnCSS} implies the inequality $\D b\geq\D b_0$, where $\D b_0\equiv \left.\D b\right|_{\Y=0}$. Therefore, for an initial CSS, a finite $\Y$ can only increase uncertainty in the frequency estimation scheme considered here. 

(ii) {\em Initial OATS}. As $\rho_0=U_\mrm{sq}\rho_{+\mvec{\hat x}}U_\mrm{sq}^\dagger$ is entangled, an exact approach is no longer viable. However, $U_\mrm{sq}$ and the effective propagators can be separated into a term that acts on qubits $n$ and $m$ in the sums of Eq.~\eq{eqnmomentsJy} and an operator acting on all other qubits. The former is evaluated and traced over exactly; the remaining expectation values are evaluated using a cumulant expansion over the system (rather than the bath), truncated to the second order~\cite{Supplement}. Neglecting higher-order terms is appropriate for 
$\q,\Psi(t)\ll1$, leading to nearly Gaussian states.
Though unwieldy, the resulting expressions will be used to obtain analytic scalings of $\D b(t)$ with $N$ for $N\gg1$.

\prlsection{Spin-boson model} To make our results quantitative, an explicit choice of noise spectra is needed. We first consider a collective spin-boson model, namely, $H_\mrm B=\hbar\sum_k \W_k a^\dagger_k a_k$ and $B(t)=2\sum_k(g_ka^\dag_k\eul{i\W_k t}+\mrm{H.c.})$, where $a_k$, $g_k$, and $\W_k$ are the annihilation operator, coupling strength, and angular frequency of bosonic mode $k$, respectively. To ease comparison with Refs.~\cite{chin2012quantum,haase2018fundamental}, we consider a continuum of bosonic modes with 
spectral density $I(\W)\equiv \al\w_c^{1-s}\W^s\eul{-\W/\w_c}$, where $\al$ is dimensionless, $\w_c$ is the cutoff frequency, and we take $s\geq 0$. Assuming that the bath is initially in its vacuum state, $\chi(t)$ and $\Psi(t)$ are readily obtained from Eqs.~\eq{eqnchi} and~\eq{eqnPsi}. From this, we calculate $\D b(t)$ for an initial CSS using Eq.~\eq{eqnCSS} with $y_0(t)\equiv 1=y(t)$ $\forall\,t$ (free evolution), and taking $\nu=T/t$, where $T$ is the {\em fixed} total available time.

In Fig.~\ref{figspinboson}(a), the uncertainty is compared with $\D b_0 (t)$. For long times, a finite $\Y$ can result in a significantly increased uncertainty. For short times, $\w_c t\ll1$, we have $\chi(t)\simeq(\chi_0 t)^2$ and $\Y(t)\simeq(\Y_0 t)^3$, with $\chi_0=\w_\mrm c{[\al\G(s+1)]^{1/2}}$ and $\Psi_0=-\w_c[\al \Gamma(s+2)/6]^{1/3}$, where $\G(x)$ is the gamma function. Upon substituting in Eq.~\eq{eqnCSS}, we find the detection time $t=t_\mrm{opt}$ that minimizes $\D b(t)$. For $N\gg1$, $t_\mrm{opt}=\chi_0^{-1}N^{-1/2}$ and $\D b_\mrm{opt}=(2\chi_0/T)^{1/2}N^{-1/4}$. This analytic scaling is intermediate between the SQL ($\D b_\mrm{opt}\propto N^{-1/2}$) and the saturation at large $N$ ($\D b_\mrm{opt}\propto \mrm{const}$) found in Ref.~\cite{dorner2012quantum} for collective Markovian noise, and coincides with the scaling obtained numerically in Ref.~\cite{frowis2014optimal} with a specific classical model of temporally correlated collective noise. 

Though $\Psi(t)$ only gives corrections of order $O(1/N)$ to $\D b(t)$ near $t=t_\mrm{opt}$, the width of the minimum in $\D b(t)$ with respect to $t$ (set by $\D b (t)\leq2\D b_\mrm{opt}$) is suppressed as $N^{-1/2}$. Experimental constraints set a minimum resolution time $t_\mrm{res}>0$; thus, even assuming perfect knowledge of the noise parameters $\alpha, s, \omega_c$ that enter $\chi_0$, it becomes harder to experimentally minimize $\D b(t)$ as $N$ increases and the dip in uncertainty shown in Fig.~\ref{figspinboson}(a) narrows.  For $t \equiv t_\mrm{opt}+t_\mrm{res}$, with $t_\mrm{res}$ fixed, $\D b(t)$ grows \emph{exponentially} with $N$ due to the term 
$\propto\cos^{2N-2}\Y(t)$ in the denominator of Eq.~\eq{eqnCSS}. This massive increase of uncertainty due to quantum noise is apparent in Fig.~\ref{figspinboson}(b), where $\D b(t)$ is seen to easily exceed $\D b_0(t)$ by orders of magnitude. Incidentally, the dips in $\D b(t)$ at long times are due to the periodicity of $\D b(t)$ with $\Psi$ ($\Psi(t)\propto t$ for $\w_c t\gg1$), and become sharper as $N$ increases.

In Fig.~\ref{figspinboson}(c), we plot $\D b(t)$ for an initial OATS with $\bt$ and $\q$ minimizing the initial uncertainty $\D J_y(0)$~\cite{kitagawa1993squeezed}. We compare the results from an exact numerical calculation of $\D b(t)$ (solid red line)~\cite{Supplement}, with those obtained from the truncated cumulant expansion over the system described earlier (dotted black line). Agreement between the two curves is excellent around $t_\mrm{opt}$, and was found to improve monotonically as $N$ increases for $1<N<1000$. For $\w_c t\ll1$ and $N\gg1$, the cumulant expansion gives $t_\mrm{opt}\simeq (4/3)^{1/6}\chi_0^{-1}N^{-5/6}$ and $\D b_\mrm{opt}\simeq (4/3)^{1/12}(2\chi_0/T)^{1/2}N^{-5/12}$. The optimal uncertainty is thus decreased by a factor $\propto N^{1/6}$ compared to an initial CSS, but is still {\em worse} than the SQL ($\propto N^{-1/2}$). As shown by the insets of Fig.~\ref{figspinboson}(c), the sharp peaks in $\D b(t)$ occuring at long times coincide with the $Q$-function of the system spiraling around the $z$ axis of the Bloch sphere, thus increasing $\D J_y$ while strongly suppressing $\moy{J_y}$. In this regime, the collective-spin state is strongly non-Gaussian, and the overall uncertainty becomes much larger than for $\Psi=0$ (dashed blue line).

\prlsection{Trapped-ion crystals} To further exemplify the adverse effects of $\Y(t)$, 
we consider the experimental setting of Ref.~\cite{gilmore2017amplitude}. Here, $N\sim 100$ ions are arranged into a 2D lattice in a Penning trap, with the electron spin in the $^2\rm{S}_{1/2}$ ground state of each $^9$Be$^+$ ion encoding a qubit. Two laser beams incident on the lattice and detuned from each other by angular frequency $\mu$ form a traveling wave, with zero-to-peak potential $U$ and wave vector $\dt k$, which couples the ions to the vibrational modes through an optical dipole force~\cite{bohnet2016quantum}. This coupling is exploited to sense the amplitude $Z_\mrm c$ of classical center-of-mass (COM) lattice motion due to a weak microwave drive applied on a trap electrode at angular frequency $\w_\mrm{rf}$. The authors estimate a single-measurement imprecision of $74$ pm, and suggest to further reduce this uncertainty by using spin-squeezed states~\cite{bohnet2016quantum} or by driving with $\w_\mrm{rf}=\mu$ near resonance with the angular frequency $\w_z$ of the COM mode. We show that quantum noise from this mode, unaccounted for in Ref.~\cite{gilmore2017amplitude}, hinders these precision improvements.

Neglecting spontaneous emission, we assume that $\w_\mrm{rf}=\mu$ is near resonance with the COM mode, with $D\equiv\w_z-\mu\ll\w_z,\mu$, but far-detuned from all other modes. Dropping terms oscillating at frequencies $\w_z+\mu,\;2\mu\gg U\dt k Z_\mrm c/\hbar$ and $\mu\gg U/\hbar$, the Hamiltonian of Eq.~\eq{eqnSB} then applies, with $b=U\dt k Z_\mrm c/\hbar$ and $B(t)=2g\,(a^\dag\eul{i\w_z t}+\mrm{H.c.})$~\cite{gilmore2017amplitude,Supplement}. Here, $a^\dag$ creates a phonon in the COM mode and $\hbar g=U\dt k\sqrt{\hbar/2MN\w_z}$, with $M$ the mass of a single ion. In addition, control of the COM mode displacement gives rise to time-dependent modulation via $y_0(t)=1-\cos D t$ and $y(t)=\cos\mu t$. Assuming that the COM mode is initially thermal, with average phonon number $\overline n$, and neglecting, again, terms oscillating at fast frequencies $\w_z+\mu$ and $2\mu$, Eqs.~\eq{eqnchi} and \eq{eqnPsi} yield $\chi(t)\simeq 8(g/ D)^2{(\overline n+\frac12)}\sin^2 (D t/2)$ and $\Y(t)\simeq2g^2\w_z t \,(1-\mrm{sinc}\,D t)/(\mu^2-\w_z^2)$.

\begin{figure}
\begin{center}
\includegraphics[width=0.47\textwidth]{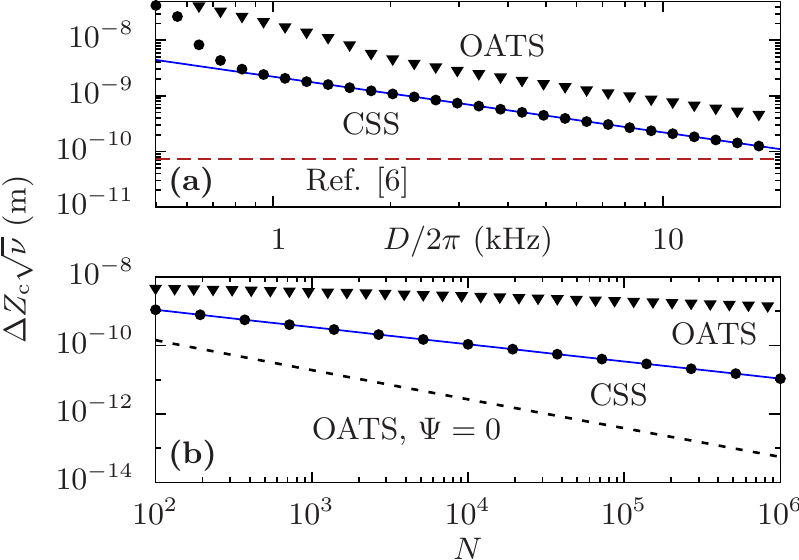}
\end{center}
\vspace{-5mm}
\caption{(Color online) 
Uncertainty in amplitude sensing with trapped ions for initial CSS vs OATS with maximal squeezing along $y$. (a) Dependence on $D$ 
for $N=100$ ions. (b) Dependence on $N$ for $D/2\pi=2$~kHz. Solid (blue) lines: analytical estimate for the initial CSS. Black dots: exact numerical optimization for the CSS. Dashed red line: uncertainty calculated in Ref.~\cite{gilmore2017amplitude} for the initial CSS. Black triangles: numerical optimization of the uncertainty from a cumulant expansion for the initial OATS. Dashed (black) line: initial OATS for $\Y=0$ within the same approach. Parameters: $\w_z/2\pi=1.57$~MHz, $U\dt k=40\times10^{-24}$~N, $M=1.50\times10^{-26}$~kg (from Ref.~\cite{gilmore2017amplitude}), and $\overline n=12.8$ resulting from a temperature of $1$~mK~\cite{sawyer2012spectroscopy}. 
\label{figIons}}
\end{figure}

Substituting the expressions of $y_0(t)$, $\chi(t)$ and $\Y(t)$ into Eq.~\eq{eqnCSS} gives $\D b(t)$ for an initial CSS. Within the regime described above, we find numerically that $t_\mrm{opt}$ occurs for $D  t_\mrm{opt} \gg 1$. For such long times, $\Y(t)$ grows linearly with $t$, while $\chi(t)$ oscillates and remains bounded by $\chi(t)\leq8(g/ D)^2(\overline n+1/2)\ll\Y(t)$, so that $\Y(t)$ provides the dominant source of uncertainty. We then approximate $\chi(t)\simeq0$ and expand the numerator and denominator of Eq.~\eq{eqnCSS} at sixth- and zeroth-order in $\Y(t)$, respectively, neglecting terms oscillating at $D$. To compare with Ref.~\cite{gilmore2017amplitude}, we optimize the single-shot detection time, considering a {\em fixed} $\nu$, and find the optimal uncertainty
$\D Z_\mrm c\simeq {U\dt k}/ [2 {\sqrt \n\,M \omega_z |D| N^{1/2}}].$	 This uncertainty is plotted in Fig.~\ref{figIons}
(solid blue lines), and shown to agree with an exact numerical optimization of Eq.~\eq{eqnCSS} (black dots) for sufficiently large $D$ and $N$. Fig.~\ref{figIons}(a) clearly shows that driving near resonance with $\w_z$ causes $\D Z_\mrm c$ to be orders of magnitude larger than estimated~\cite{gilmore2017amplitude} by neglecting correlated quantum noise (dashed red line).

Finally, we evaluate the uncertainty in amplitude sensing with an initial OATS. Taking initial values of $\bt$ and $\q$ that minimize initial uncertainty along $y$, we numerically optimize $t_\mrm{opt}$, using again a truncated cumulant expansion over the system. The black triangles in Fig.~\ref{figIons} show that rather than improving precision, this initial OATS leads to an uncertainty that is {\em larger} and suppressed more slowly with $N$ than for an initial CSS (a numerical fit gives $\D b\propto N^{-1/6}$). Thus, not only does this correlated quantum noise prevent the realization of the superclassical scaling $\D b\propto N^{-5/6}$ that would arise for $\Y=0$ (dashed black line in Fig.~\ref{figIons}(b)); but, in fact, the collective-spin state becomes ``anti-squeezed'' along the $y$ axis, making the scaling even worse than the SQL.

\prlsection{Discussion} Interestingly, for collective noise as we consider here, the reduced state of the system can be  
written as $\rho_\mrm S(t)=U_\Y(t) (\left.\rho_\mrm S(t)\right|_{\Y=0} ) U_\Y^\dag(t)$, with $U_\Y(t)\equiv \exp[-i\Y(t)J_z^2]$~\cite{Next}. The quantum Fisher information being invariant under unitary transformations that do not depend on $b$~\cite{braunstein1996generalized}, there always exists an {\em optimal} measurement that cancels the effect of $\Y(t)$ in principle. However, not only is this measurement highly non-local in general, 
but it requires precise knowledge of $\Y(t)$. This makes it far more challenging from an implementation standpoint.

In summary, we showed that spatio-temporally correlated quantum noise with frequency asymmetric spectra can generate unwanted entanglement of the sensing system that hinders superclassical precision scaling in Ramsey interferometry. Beside amplitude sensing 
with trapped ions, such noise sources arise naturally in a variety of other platforms -- notably, superconducting qubits \cite{nonclassical,rigetti2012superconducting}, nitrogen-vacancy centers \cite{astner2017nv}, or spin qubits in semiconductors 
\cite{petta2018}, in which qubit coupling to a common microwave cavity yields correlated photon shot noise. Our result is also directly relevant to ultrasensitive magnetometry and atomic clocks, 
as both fields are moving toward larger ensembles of entangled probes to reduce uncertainty below the shot-noise limit~\cite{magnetometry,clocks}. This highlights the need for accurate characterization of quantum noise~\cite{paz2017multiqubit}, 
which may allow for counteracting unwanted entanglement through appropriate initialization, measurement, 
or dynamical control \cite{paz2016dynamical}.

It is a pleasure to thank Sandeep Mavadia and Jun Ye for useful discussions.
F. B. acknowledges support from the {\em Fonds de Recherche du Qu\'ebec -- Nature et Technologies}. 
Partial support from the the US Army Research Office under Contract W911NF-12-R-0012 is also 
gratefully acknowledged.


\bibliographystyle{apsrev4-1}
\bibliography{paper}

\begin{thebibliography}{45}
\bibitem [{Cla()}]{Classical}%
  \BibitemOpen
  \href@noop {} {\ }\bibinfo {note} {Though $\Y(t)\neq0$ implies non-commuting
  (quantum) noise, note that the converse need not be true: for instance,
  $\Y(t)=0$ for $\delta$-correlated (white) quantum noise.}\BibitemShut {Stop}%
\end{thebibliography}

\end{document}


\title{Supplemental Material for ``Ramsey Interferometry in Correlated Quantum Noise Environments''}
\author{F\'elix Beaudoin}
\affiliation{$\mbox{Department of Physics and Astronomy, Dartmouth College, 6127 Wilder Laboratory, 
Hanover, New Hampshire 03755, USA}$}
\author{Leigh M. Norris}
\affiliation{$\mbox{Department of Physics and Astronomy, Dartmouth College, 6127 Wilder Laboratory, 
Hanover, New Hampshire 03755, USA}$}
\author{Lorenza Viola}
\affiliation{$\mbox{Department of Physics and Astronomy, Dartmouth College, 6127 Wilder Laboratory, 
Hanover, New Hampshire 03755, USA}$}

\date{\today}
\maketitle

\onecolumngrid 

\vspace*{-10mm}

\setcounter{figure}{0}
\setcounter{equation}{0}
\setcounter{section}{0}

\section{Uncertainty for initial coherent spin states and one-axis-twisted states}

In this Section, we discuss the basic steps involved in the calculation of the uncertainty of a frequency estimate under the Hamiltonian $H_\mrm{SB}(t)$, defined in Eq.~(1) of the main text.
We begin by revisiting an exact result, established in Ref.~\cite{paz2017multiqubit} [Sec. II.D. and Appendix A therein], 
which will be used extensively.
Let $O$ be an \emph{invertible} operator acting on the system only. Then, assuming an initial factorizable joint state $\rho_\mrm{SB}(0)\equiv \rho_0\otimes\rho_B$ of the system and bath, we can express its time-dependent expectation value as
\begin{align}
 \mean{O(t)}&= \Tr_\mrm S\left[\bmean{\mathcal T_+\exp\left[-\frac i\hbar\int_{-t}^tds\,H_O(s)\right]}_\mrm B\rho_0 O\right],	\label{eqnIdentity1}
\end{align}
where $\mathcal T_+$ is the time-ordering operator and, for an arbitrary operator $A$ acting in general on both the system and the bath, the partial trace $\mean{A}_\mrm B\equiv \Tr_\mrm B[A\,\rho_\mrm B]$ may be taken to define a ``generalized average,'' in the sense of Kubo~\cite{kubo1962generalized}. In Eq.~\eq{eqnIdentity1}, we have also introduced an operator-dependent effective Hamiltonian 
\begin{align}
H_O(s) \equiv \left\{
\begin{array}{ll}
 -O^{-1}H_\mrm{SB}(t-s)O\hspace{1cm}	&	\mbox{for }0<s\leq t,\\
H_\mrm{SB}(t+s)		&	\mbox{for }-t\leq s<0.
\end{array}
\right.	\label{eqnEffectiveHamiltonian}
\end{align}

Assuming that $O$ is \emph{dephasing-preserving}, in the sense that $O^{-1}\s_n^z O=\sum_\ell V_{n\ell}\s^z_\ell\;\forall\;n$, where $V_{n\ell}\in\mathbb C$, and that noise is \emph{Gaussian}, a generalized cumulant expansion truncates exactly to the second order, namely, 
\begin{align}
\bmean{\mathcal T_+\exp\left[-\frac i\hbar\int_{-t}^tds\,H_O(s)\right]}_\mrm B=\exp\left[-i\,\mathcal C^{(1)}_O(t)-\mathcal C^{(2)}_O(t)/2\right],
\label{eqnExpansion}
\end{align}
where $\mathcal C^{(k)}_O(t)$ is the $k$-th order generalized cumulant~\cite{kubo1962generalized} for operator $O$. Explicitly, 
\begin{align}
 \mathcal C_O^{(1)}(t)&=\frac1\hbar\int_{-t}^tds\mean{H_O(s)},	\label{eqnC1O}\\
 \frac{\mathcal C_O^{(2)}(t)}2&= \frac1{\hbar^2}\int_{-t}^t ds_1\int_{-t}^{s_1}ds_2\,\mean{H_O(s_1)H_O(s_2)}_\mrm B-\frac1{2\hbar^2}\int_{-t}^tds_1\int_{-t}^t ds_2\mean{H_O(s_1)}_\mrm B\mean{H_O(s_2)}_\mrm B.	\label{eqnC2O}
\end{align}

\subsection{Expectation values of collective spin operators for arbitrary initial states 	
\label{eqnExpectation}}

We now proceed to evaluate the following expectation values of collective-spin operators:
\begin{equation}
 \mean{J_y(t)}=\sum_{n=1}^N \frac{\mean{\s^y_n(t)}}2,
 \hspace{1cm}
 \mean{J_y^2(t)}=\frac N4 +\frac14\sum_{\underdot n,\underdot m\neq n}\mean{\s^y_n\s^y_m(t)},	
 \label{eqnCollectiveSpin}
\end{equation}
where the lower dot marks indices over which sums are performed, in cases that could otherwise be ambiguous.

\vspace{3mm}
Consider first the expectation value $\mean{\s^y_n(t)}$. Since $\s^y_n$ is an invertible operator, $\mean{\s^y_n(t)}$ is given by Eq.~\eq{eqnIdentity1}. In addition, $\s^y_n$ is dephasing-preserving: $\s^y_n\s^z_n\s^y_n=-\s^z_n$. We can thus directly use the cumulant expansion in Eq.~\eq{eqnExpansion} to evaluate the expectation value over the bath.
Assuming, as in the main text, that noise is zero-mean and stationary, that is, $\mean{B_n(s)}=0\;\forall\;n,s$, and 
$\mean{B_n(s_1)B_n(s_2)}_\mrm B=\mean{B_n(s_1-s_2)B_n(0)}_\mrm B$, the first and second cumulants are given by
\begin{align}
 \mathcal C_{\s^y_n}^{(1)}(t)=\vf(t)\s^z_n,
 \hspace{1cm}
 \frac12{\mathcal C_{\s^y_n}^{(2)}(t)}=\frac{\chi_{nn}(t)}2 + i\sum_{\underdot\ell\neq n}\Y_{n\ell}(t)\s^z_n\s^z_\ell,	\label{eqnsyCumulantsExplicit}
\end{align}
where $\vf(t)$, $\chi_{nm}(t)$ and $\Y_{nm}(t)$ are defined in the main text. Substituting, in turn, Eqs.~\eq{eqnsyCumulantsExplicit} into Eq.~\eq{eqnExpansion}, Eq.~\eq{eqnExpansion} into Eq.~\eq{eqnIdentity1} with $O\rightarrow\s^y_n$, and Eq.~\eq{eqnIdentity1} into Eq.~\eq{eqnCollectiveSpin}, then yields the expression for $\mean{J_y(t)}$ quoted in Eq.~(3) of the main text.

\vspace{3mm}
Since $\s^y_n\s^y_m$ is also invertible and dephasing-preserving, we can follow similar steps to the above to evaluate 
$\mean{\s^y_n\s^y_m(t)}$. Under the same assumptions, the relevant cumulants are
\begin{align}
 \mathcal C_{\s^y_n\s^y_m}^{(1)}(t)&=\vf(t)(\s^z_n+\s^z_m),	\label{eqnsy2C1}\\
 \frac12{\mathcal C_{\s^y_n\s^y_m}^{(2)}(t)}
 &= \frac{\chi_{nn}(t)+\chi_{mm}(t)}2+\chi_{nm}(t)\s^z_n\s^z_m
  +i\!\!\!\sum_{\underdot\ell\neq n,m}\big[\Y_{n\ell}(t)\s^z_n\s^z_\ell + \Y_{m\ell}(t)\s^z_m\s^z_\ell\big].\label{eqnsy2C2}
\end{align}
Proceeding as above, we find the expression for $\mean{J_y^2(t)}$ quoted in Eq.~(3) of the main text.

\subsection{Uncertainty for an initial coherent spin state or one-axis twisted state	
\label{secCSSOATS}}

The above general results can now be specialized to the relevant cases of an initial CSS or OATS, by performing the traces over the system (i.e., the $N$ qubits) appearing in Eq.~(3) of the main text. The resulting expectation values then directly yield the uncertainty $\D b(t)$ through Eq.~(2) of the main text.

\subsubsection{Evaluation of $\mean{J_y(t)}$}

We first evaluate $\mean{J_y(t)}$ for an initial OATS, which we express in the form
\begin{align}
\rho_0=\ket S\bra S,
\hspace{1cm}
\ket S\equiv \exp(-i\bt J_x)\exp(-i\q J_z^2/2)\,\ket{+}^{\otimes N}.	
\label{eqndefS}
\end{align}
In Eq.~(3) of the main text, $\mean{J_y(t)}$ is given by a sum over qubits labeled by $n$. The key to our approach is to treat qubit $n$ and qubits $\ell\neq n$ separately. To do so, we introduce new collective spin operators
\begin{align}
J_{\neq n}^\al\equiv\sum_{\underdot\ell\neq n}\frac{\s^\al_\ell}2,
\hspace{1cm}\al\;\in\;\{x,y,z\},	
\label{eqnCollectiveSpindiffn}
\end{align}
that do not involve qubit $n$. Using $J_\al=J^\al_{\neq n}+\s^\al_n/2$, we write the OATS as
\begin{align}
 \ket S&=\eul{-i\q/8}\exp(-i\bt J^x_{\neq n})\exp\left[-i\q(J^z_{\neq n})^2/2\right]\exp(-i\bt\s^x_n/2)\exp(-i\q J^z_{\neq n}\s^z_n/4)\ket{+}_n\ket{+}_{\neq n},	\label{eqnSseparaten}
\end{align}
where $\ket+_{\neq n}\equiv\bigotimes_{\underdot\ell\neq n}\ket+_\ell$ is a CSS excluding qubit $n$. 
We also write the effective propagator as
\begin{align}
 \eul{-i\F_n(t)}=\exp\left\{-i\left[\vf(t)+\y_{\neq n}(t)\right]\s^z_n\right\},
 \hspace{1cm}
 \y_{\neq n}(t)\equiv\sum_{\underdot\ell\neq n}\Y_{n\ell}(t)\s^z_\ell\,.	
\label{eqnpropseparaten}
\end{align}
We then substitute Eqs.~\eq{eqnSseparaten} and \eq{eqnpropseparaten} into Eq.~(3) of the main text for $\mean{J_y(t)}$, and evaluate all expectation values of operators for qubit $n$ with respect to $\ket +_n$ exactly. Summing over all qubits, this results in
\begin{align}
 \mean{J_y(t)}&=\sum_n\frac{\eul{-\chi_{nn}(t)/2}}4
  \left[
    \frac{\eul{-i\vf(t)}}2\sin\bt\left(\mathscr E^{\neq n}_{--+}-\mathscr E^{\neq n}_{+--}\right)\right.+
    i\eul{-i\vf(t)}\sin^2\left(\frac\bt2\right) \;\mathscr E^{\neq n}_{+-+}-\left.
    i\eul{i\vf(t)}\cos^2\left(\frac\bt2\right) \;\mathscr E^{\neq n}_{+++}+\mrm{c.c.} \right],	
 \label{eqnsyoverlaps}
\end{align}
where $\mathscr E^{\neq n}_{s_1s_2s_3}$ represents the remaining expectation values relevant to all qubits $\ell\neq n$. Explicitly,
\begin{align}
 \mathscr E^{\neq n}_{s_1s_2s_3}&=\bra{S_{\neq n}}\exp\left(s_1\,i\q J^z_{\neq n}/2\right)\exp\left[s_2\,i\widetilde\y_{\neq n}(t)\right]\exp\left(s_3\,i\q J^z_{\neq n}/2\right)\ket{S_{\neq n}},
 \hspace{1cm}
 s_1,s_2,s_3\;\in\;\{+1,-1\},	
 \label{eqnEn}
\end{align}
with
\begin{align}
\widetilde\y_{\neq n}(t) \equiv \exp(i\bt J^x_{\neq n})\y_{\neq n}(t)\exp(-i\bt J^x_{\neq n}),
\hspace{1cm}
\ket{S_{\neq n}}\equiv \exp\left[-i\q(J^z_{\neq n})^2/2\right]\ket{+}_{\neq n}.	
\label{eqndefSn}
\end{align}
So far, no approximation has been made. 
In addition, Eq.~\eq{eqnsyoverlaps} no longer involves any operator involving qubit $n$. In what follows, we will treat the remaining expectation values for qubits $\ell\neq n$ first exactly for the initial CSS, and then approximately for the OATS.

\vspace{3mm}
\textbf{$\bullet$ Initial CSS.}
For an initial CSS, we take $\q=\bt=0$ in Eqs.~\eq{eqnsyoverlaps} to \eq{eqndefSn}. Though this gives a general expression valid for arbitrary spatial correlations, here we focus on collective noise for simplicity, $B_n(t)=B(t)\;\forall\;n,t$, resulting in $\chi_{nm}(t)=\chi(t)$ and $\Y_{nm}(t)=\Y(t),\;\forall\;n,m,t$. We then find
\begin{align}
 \mean{J_y(t)}=\frac N2\eul{-\chi(t)/2}\sin[\vf(t)]\cos^{N-1}\left[\Y(t)\right].
\end{align}

\textbf{$\bullet$ Initial OATS.} For an initial OATS, because $\ket{S_{\neq n}}$ is entangled, we cannot find exact closed-form expressions for $\mathscr E^{\neq n}_{s_1s_2s_3}$. However, in Eq.~\eq{eqnEn} each $\mathscr E^{\neq n}_{s_1s_2s_3}$ takes the form of an expectation value with respect to $\ket{S_{\neq n}}$. This expectation value defines a generalized average, in the sense of Kubo~\cite{kubo1962generalized}. Therefore, we can perform a generalized cumulant expansion for each $\mathscr E^{\neq n}_{s_1s_2s_3}$. Truncating this expansion to second order and expressing cumulants in terms of moments then gives
\begin{align}
 &\mathscr E^{\neq n}_{s_1s_2s_3}\simeq\exp\left\{i(s_1+s_3)\frac\q2\bmean{J^z_{\neq n}}_{\neq n}+is_2\bmean{\widetilde\y_{\neq n}(t)}_{\neq n}
 -\frac{\q^2}4(1+s_1s_3)\left(\bmean{(J^z_{\neq n})^2}_{\neq n}-\bmean{J^z_{\neq n}}_{\neq n}^2\right)\right.  \label{eqnCumulantsJy}\\
 &\qquad\left.-\frac12\left(\bmean{\widetilde\y_{\neq n}(t)^2}_{\neq n}-\bmean{\widetilde\y_{\neq n}(t)}_{\neq n}^2\right)
  - \frac\q2\left[
    s_1s_2\bmean{J^z_{\neq n}\widetilde\y_{\neq n}(t)}_{\neq n}+s_2s_3\bmean{\widetilde\y_{\neq n}(t)J^z_{\neq n}}_{\neq n}
    -(s_1s_2+s_2s_3)	\bmean{J^z_{\neq n}}_{\neq n}\bmean{\widetilde\y_{\neq n}(t)}_{\neq n}
  \right]
 \right\},	\notag
\end{align}
where $\mean{A}_{\neq n}\equiv\bra{S_{\neq n}}A\ket{S_{\neq n}}$ for an arbitrary operator $A$. Evaluating these expectation values, and substituting the resulting $\mathscr E^{\neq n}_{s_1s_2s_3}$ in Eq.~\eq{eqnsyoverlaps} then yields $\mean{J_y(t)}$ for arbitrary spatial correlations. Assuming collective noise for simplicity, we then find
\begin{align}
 \mean{J_y(t)}&\simeq\frac N2\eul{-\chi(t)/2}\sin[\vf(t)]\exp[-\mean{\widetilde\y_{\neq n}(t)^2}_{\neq n}/2]\notag\\
  &\hspace*{-5mm}\qquad\left\{
    \eul{-\q^2(N-1)/8}
    \left[
      \cos^2\left(\frac\bt2\right)\exp\left[-\q\,\mrm{Re}\mean{J^z_{\neq n}\widetilde\y_{\neq n}(t)}_{\neq n}\right]
      +\sin^2\left(\frac\bt2\right)\exp\left[\q\,\mrm{Re}\mean{J^z_{\neq n}\widetilde\y_{\neq n}(t)}_{\neq n}\right]\right]\right.\notag\\
  &\hspace*{-5mm}\qquad+\left.\sin\bt\sin\left[(N-1)\frac\q2\sin\bt\cos^{N-2}\left(\frac\q2\right)\Y(t)\right]\right\},	\label{eqnJyOATS}
\end{align}
where
\begin{align}
 \mean{\widetilde\y_{\neq n}(t)^2}_{\neq n}&=(N-1)\Y(t)^2\left\{
  1+(N-2)\left[\frac12\sin^2\bt\left(1-\cos^{N-3}\q\right)+\sin(2\bt)\sin\left(\frac\q2\right)\cos^{N-3}\left(\frac\q2\right)\right]
 \right\},\\
 \mrm{Re}\mean{J^z_{\neq n}\widetilde\y_{\neq n}(t)}_{\neq n}&=(N-1)\frac{\Y(t)}2\left[
    \cos\bt+(N-2)\sin\bt\sin\left(\frac\q2\right)\cos^{N-3}\left(\frac\q2\right)
 \right].	\label{eqnR1OATS}
\end{align}
Note that the collective spin state is approximately Gaussian-distributed when the non-linearities entering the problem are small, $\q,\Y(t)\ll1$. The Gaussian approximation made in Eq.~\eq{eqnCumulantsJy} by truncating the cumulant expansion at second order is then appropriate, 
with the leading non-Gaussian correction arising from fourth-order cumulants.
Evaluating these cumulants, we have verified that they result in a slightly improved agreement with the exact numerical solution for collective noise (see Section~\ref{secExact}, below). These corrections, however, are negligible in the specific scenarios studied in the main text (in particular, for $t \lesssim t_{\text{opt}}$).

\subsubsection{Evaluation of $\mean{J_y^2(t)}$}

Evaluation of $\mean{J_y^2(t)}$ proceeds along similar steps as above. To treat qubits $n$ and $m$ separately from qubits $\ell\neq n,m$, we introduce a new collective spin operator
\begin{align}
 J^\al_{\neq nm}\equiv\sum_{\underdot\ell\neq nm}\frac{\s^\al_\ell}2,
 \hspace{1cm}
 \al\;\in\;\{x,y,z\}.
\end{align}
Using $J_\al=J^\al_{\neq nm}+\s^\al_n/2+\s^\al_m/2$, we then write the initial OATS as
\begin{align}
 \ket S&=\eul{-i\q/4}\exp[-i\bt J^x_{\neq nm}]\exp[-i\q(J^z_{\neq nm})^2/2]\notag\\
  &\qquad\times\exp[-i\bt(\s^x_n+\s^x_m)/2]\exp[-i\q J^z_{\neq nm}(\s^z_n+\s^z_m)/2]\exp(-i\q\s^z_n\s^z_m/4)\ket+_n\ket+_m\ket+_{\neq nm},	\label{eqnSseparatenm}
\end{align}
where $\ket+_{\neq nm}\equiv\prod_{\ell\neq nm}\ket+_\ell$. We also write the effective propagator as
\begin{align}
 \eul{-i\F_{nm}(t)}&=\exp\left[-i\vf(t)(\s^z_n+\s^z_m)-\chi_{nm}(t)\s^z_n\s^z_m-i\y_{\neq nm}(t)\s^z_n-i\y_{\neq mn}(t)\s^z_m\right],
 \hspace{0.7cm}
 \y_{\neq nm}(t)\equiv\sum_{\underdot\ell\neq nm}\Y_{n\ell}(t)\s^z_\ell.	\label{eqneffectivesynsym}
\end{align}
We then substitute Eqs.~\eq{eqnSseparatenm} and \eq{eqneffectivesynsym} into Eq.~(3) of the main text for 
$\mean{J_y^2(t)}$ and evaluate all expectation values of operators for qubits $n$ and $m$ with respect to $\ket+_n\ket+_m$ exactly. In this way, we are left with expectation values of operators acting on all other qubits $\ell \neq n,m$, taken with respect to the state $\ket{S_{\neq nm}}\equiv\exp\left[-i\q(J^z_{\neq nm})^2/2\right]\ket+_{\neq nm}$. Again, these expectation values are evaluated exactly for an initial CSS ($\q=\bt=0$), and approximately for an initial OATS.

\vspace{3mm}
\textbf{$\bullet$ Initial CSS.} For an initial CSS, and assuming collective noise, we find
\begin{align}
 \bmean{J_y^2(t)}=\frac N8\left\{N+1-(N-1)\cos[2\vf(t)]\eul{-2\chi(t)}\cos^{N-2}\left[2\Y(t)\right]\right\}.
\end{align}

\textbf{$\bullet$ Initial OATS.} For an initial OATS, we perform a cumulant expansion to evaluate expectation values with respect to $\ket{S_{\neq nm}}$. Truncating to the second order, and again assuming collective noise for simplicity yields 
\begin{align}
 \mean{J_y^2(t)}&\simeq\frac N4+\frac{N(N-1)}2\left\{\Upsilon_1(t)-\cos[2\vf(t)]\eul{-2\chi(t)}\Upsilon_2(t)\right\} ,
\end{align}
where
\begin{align}
 \Upsilon_1(t)&=\frac{1+\cos^2\bt}8-\frac14\eul{-(N-2)\q^2/8}\sin(2\bt)\sin\left(\frac\q2\right)+\frac18\eul{-(N-2)\q^2/2}\sin^2\bt,	\label{eqnUps1}\\
 \Upsilon_2(t)&=\!\left\{
    \frac14\sin^2\bt
     +\frac14\left[\cos^4\left(\frac\bt2\right)\eul{-4\q \z_1}
     +\sin^4\left(\frac\bt2\right)\eul{4\q \z_1}\right]
     \eul{-(N-2)\q^2/2}
     -\frac12\cos^2\left(\frac\bt2\right)\sin^2\left(\frac\bt2\right)\cos\left(4\q \z_2\right)\right.	 \label{eqnUps2} \\
 &+\left.\frac12\sin\bt\left[\cos^2\left(\frac\bt2\right)\sin\left(\frac\q2\left(1-4\z_2\right)\right)\eul{-2\q \z_1} \!
    -\sin^2\left(\frac\bt2\right)\sin\left(\frac\q2\left(1+4\z_2\right)\right)\eul{2\q \z_1}\right]\eul{-(N-2)\q^2/8}
    \!\right\}\exp\left[\!-2\bmean{\widetilde\y_{\neq nm}(t)^2}_{\neq nm}\right].	\notag
\end{align}
In Eq.~\eq{eqnUps2}, we have introduced the operator $\widetilde\y_{\neq nm}(t)=\exp[i\bt J^x_{\neq nm}]\y_{\neq nm}(t)\exp[-i\bt J^x_{\neq nm}]$, along with moments and covariances taken with respect to the state $\ket{S_{\neq nm}}$:
\begin{align}
 &\z_1\equiv\mrm{Re}\mean{J^z_{\neq nm}\widetilde \y_{\neq nm}(t)}_{\neq nm}=\frac12(N-2)\Y(t)\left[\cos\bt+(N-3)\sin\bt\sin\left(\frac\q2\right)\cos^{N-4}\left(\frac\q2\right)\right],\\
 &\z_2\equiv\mrm{Im}\mean{J^z_{\neq nm}\widetilde \y_{\neq nm}(t)}_{\neq nm}=-\frac12(N-2)\Y(t)\sin\bt\cos^{N-3}\left(\frac\q2\right),\\
 &\bmean{\widetilde\y_{\neq nm}(t)^2}_{\neq nm} = (N-2)\Y^2(t)\left\{
  1+(N-3)\left[\frac12\sin^2\bt\left(1-\cos^{N-4}\q\right)+\sin(2\bt)\sin\left(\frac\q2\right)\cos^{N-4}\left(\frac\q2\right)\right]
  \right\}.
\end{align}

\subsection{Numerical calculations for collective spin states	\label{secExact}}

\subsubsection{Reduced density matrix}

Using the Hamiltonian $H_\mrm{SB}(t)$ of Eq.~(1) of the main text and assuming collective noise,  $B_n(t)=B(t)\;\forall\;n$, the time-evolved reduced density operator for the system is
\begin{align}
 \rho_\mrm S(t)
 =\Tr_\mrm B \left\{
  \mathcal T_+\exp\left[- \frac{i}{\hbar} \int_0^tds\,H_\mrm{SB}(s)\right]
  (\rho_0\otimes\rho_\mrm B)
  \mathcal T_-\exp\left[\frac{i}{\hbar}\int_0^tds\,H_\mrm{SB}(s)\right] \right\},
 \hspace{1cm}
 H_\mrm{SB}(t)=\hbar \widetilde B(t)J_z,
 \label{eqnDefReduced}
\end{align}
with $\widetilde B(t)\equiv y_0(t) b+y(t)B(t)$ and where $\mathcal T_-$ is the anti-time-ordering operator, which sorts time-dependent operators in antichronological order. For collective noise, $H_\mrm{SB}(t)$ preserves the total spin angular momentum 
$\mvec J^2$, with $\mvec J=(J_x,J_y,J_z)$. Thus, the resulting evolution will not couple states belonging to different irreducible representations of the rotation group. In particular, 
permutationally invariant initial pure states, such as the CSS or the OATS, belong to the fully symmetric (highest weight) irreducible representation, labeled by $J=N/2$ (in units of $\hbar$).
We thus write the corresponding $\rho_0$ in the basis $\{\ket{J,M}\}$ of eigenstates of $J_z$: $J_z\ket{J,M}= M\ket{J,M}$. 
Explicitly, we have
\begin{align}
 \rho_0=\sum_{M_1M_2}\rho_{M_1M_2}\ket{J,M_1}\bra{J,M_2},
 \hspace{1cm}
 \rho_{M_1M_2}\equiv\bra{J,M_1}\rho_0\ket{J,M_2}.
 \label{eqnRhoEigenbasis}
\end{align} 
Substituting Eq.~\eq{eqnRhoEigenbasis} into Eq.~\eq{eqnDefReduced} then gives
\begin{align}
 \rho_\mrm S(t)&=\sum_{M_1M_2}\rho_{M_1M_2}\bmean{\mathcal T_-\exp\left[i\int_0^tds\,\widetilde B(s)M_2\right]\mathcal T_+\exp\left[-i\int_0^tds\,\widetilde B(s)M_1\right]}_\mrm B\ket{J,M_1}\bra{J,M_2}\notag\\
 &=\sum_{M_1M_2}\rho_{M_1M_2}\bmean{\mathcal T_+\exp\left[-\frac i\hbar\int_{-t}^tdsH_{M_1M_2}(s)\right]}_\mrm 
 B\ket{J,M_1}\bra{J,M_2}	,
 \label{eqnReduced} 
\end{align}
where we have introduced the effective Hamiltonian
\begin{align}
 H_{M_1M_2}(s) \equiv \left\{
  \begin{array}{ll}
   -\hbar\widetilde B(t-s)M_2	\hspace{1cm}\mbox{for }0<s\leq t,\\
   \hbar\widetilde B(t+s)M_1	\hspace{1cm}\mbox{for }-t\leq s<0,
  \end{array}
 \right.
\end{align}
As above, we evaluate the average $\mean\cdot_\mrm B$ in Eq.~\eq{eqnReduced} from the generalized cumulant expansion using Eqs.~\eq{eqnExpansion} to \eq{eqnC2O} for Gaussian noise. Assuming, as before, stationary noise with zero mean, we then find
\begin{align}
 \rho_\mrm S(t)=\sum_{M_1M_2}\rho_{M_1M_2}\exp\left[-i\vf(t)(M_1-M_2)-i\Y(t)(M_1^2-M_2^2)-\chi(t)(M_1-M_2)^2\right]\ket{J,M_1}\bra{J,M_2}.	
 \label{eqnrhos}
\end{align}
Given $\vf(t)$, $\chi(t)$, and $\Y(t)$ from a specific control setting and noise model, $\rho_\mrm S(t)$ is readily evaluated numerically from Eq.~\eq{eqnrhos} and the matrix elements $\rho_{M_1M_2}$ for an initial CSS or OATS.

\subsubsection{Uncertainty in frequency estimation and Husimi $Q$-functions}

The reduced density matrix $\rho_\mrm S(t)$ given in Eq.~\eq{eqnrhos} allows us to produce the numerical calculations of $\D b(t)$ and of the $Q$ functions shown in Fig.~1(c) of the main text. From Eq.~(2) of the main text, $\D b(t)$ is determined by $\mean{J_y(t)}$, $\mean{J_y^2(t)}$ and $\partial \mean{J_y(t)}/\partial b$. $\mean{J_y(t)}$ and $\mean{J_y^2(t)}$ are evaluated numerically from $\mean{O(t)}=\Tr_\mrm S[O\rho_\mrm S(t)]$; likewise, $\partial \mean{J_y(t)}/\partial b$ is given by
\begin{align}
\hspace*{-4mm} \del{\mean{J_y(t)}}{b}&=\left[\int_0^tds\,y_0(s)\right]\Tr_\mrm S\left[J_y\del{\rho_\mrm S(t)}{\vf}\right],	\label{eqnDerivativeJy}\\
\hspace*{-4mm}\del{\rho_\mrm S(t)}{\vf}&=-i\sum_{M_1M_2}(M_1-M_2)\rho_{M_1M_2}\exp\left[-i\vf(t)(M_1-M_2)-i\Y(t)(M_1^2-M_2^2)-\chi(t)(m_1-m_2)^2\right]\ket{J,M_1}\bra{J,M_2}.
  \label{eqnDerivativerhos}
\end{align}
Finally, the $Q$-functions shown in Fig.~1 of the main text are defined by
$ Q(\vartheta,\g) \equiv \bra{\vartheta,\g}\rho_\mrm S(t)\ket{\vartheta,\g},$
where $\rho_\mrm S(t)$ is evaluated from Eq.~\eq{eqnrhos} and where $\ket{\vartheta,\gamma}$ is an arbitrary CSS with polar and azimuthal angles $\vartheta$ and $\g$, respectively, that is, 
\begin{align}
  \ket{\vartheta,\g}=\exp[-i\vartheta(J_x\sin\g-J_y\cos\g)]\ket{J=N/2,M=-N/2}.	
  \label{eqndefCSS}
\end{align}

\section{Quantum noise in amplitude sensing with trapped ions}

In this Section, we show how the trapped-ion setting of Ref.~\cite{gilmore2017amplitude} is captured by the model Hamiltonian of Eq.~(1) of the main text over time scales of experimental relevance. In this experiment,  the goal is to sense the amplitude $Z_\mrm c$ of coherent oscillations of the COM mode of a lattice of $N$ ions confined in a Penning trap. This is achieved by coupling the qubits encoded by the electron spin of the ions to the vibrational modes through the optical-dipole force (ODF) produced by a pair of laser beam intersecting at the lattice. To demonstrate the approach, oscillations of the COM mode are driven by a uniform rf electric field along the axial direction $z$. Deep in the Lamb-Dicke confinement regime, this setting is described by the Hamiltonian
$ H(t)= H_\mrm{vib}+H_\mrm{rf}(t)+H_\mrm{ODF}(t),	$
where the three contributions have the following explicit form:
\begin{align}
 &H_\mrm{vib}=\hbar\sum_k\w_k a^\dag_k a_k,	\label{eqnHvib}\\
 &H_\mrm{rf}(t)=\ve\cos(\w_\mrm{rf}t+\dt)\sum_n z_n,	\label{eqnHrf}\\
 &H_\mrm{ODF}(t)=U\dt k\cos(\m t-\f)\sum_n z_n\s^z_n-  U\sin(\m t-\f) \sum_n\s^z_n.	
 \label{eqnHODF}
\end{align}
In the above equations, $a_k$ is the bosonic operator that annihilates the vibrational mode $k$ of the lattice with angular frequency $\w_k$. We have also introduced $\w_\mrm{rf}$ and $\dt$, the frequency and phase of the applied rf field, and $\ve=-qE$, with $q$ the charge of an ion and $E$ the amplitude of the rf electric field along $z$. In addition, $U$, $\dt k$, and $\phi$ are, respectively, the zero-to-peak potential, wave vector, and phase of the 1D traveling wave produced by the pair of laser beams with frequency difference $\mu$. Finally, $z_n$ is the position operator for ion $n$, and is given by~\cite{sawyer2012spectroscopy}
\begin{align}
 z_n=\sum_{k=1}^N v_{nk}\sqrt{\frac\hbar{2M\w_k}}(a_k+a^\dag_k),
\end{align}
where $M$ is the mass of an individual ion and $v_{nk}$ is the displacement amplitude of ion $n$ in vibrational mode $k$.

The COM mode, here labeled by $k=1$, is defined by $v_{n1}=1/\sqrt{N}\;\forall\;n$. Orthonormality of the vibrational modes then leads to $\sum_n v_{nk}=\dt_{k1}\sqrt N$. Because of this last equality, $\sum_n z_n=Z_1(a_1+a_1^\dag)$, where $Z_1\equiv\sqrt{N\hbar/2M\w_z}$, with $\w_z$ the frequency of the COM mode. Substituting this into Eq.~\eq{eqnHrf} shows that $H_\mrm{rf}(t)$ produces coherent oscillations of the COM mode only. This is a consequence of the assumption of a uniform rf electric field. We account for the dynamics of the COM mode by transforming $H(t)$ into a displaced frame with
$ H_R(t)=R^\dag(t) H(t)R(t)-i\hbar R^\dag(t)\dot R(t),$
$R(t)=\exp\left[\al_1(t)a^\dag_1-\al^\ast_1(t)a_1\right] ,$
where the unitary transformation $R(t)$ generates time-dependent displacements $\al_1(t)$ of the COM mode. Because $R(t)$ commutes with any qubit operator, expectation values $\mean{J_y(t)}$ and $\mean{J_y^2(t)}$ evaluated in the main text are unaffected by this transformation. Taking
\begin{align}
 \al_1(t)&=\frac{\ve Z_1}{\hbar(\w^2_\mrm{rf}-\w_z^2)}\Big\{\left[\w_z\cos(\w_\mrm{rf} t+\dt)+\w_\mrm{rf}\sin(\w_z t)\sin\dt-\w_z\cos(\w_z t)\cos \dt\right]	\notag\\
 &\hspace*{-4mm}\qquad-i\left[\w_\mrm{rf}\sin(\w_\mrm{rf}t+\dt)-\w_\mrm{rf}\cos(\w_z t)\sin\dt-\w_z\sin(\w_z t)\cos\dt\right]\Big\},
    \label{eqnalpha1}
\end{align}
then results in
$ H_R(t)=H_\mrm{vib}+H_\mrm{ODF}(t)+V_R(t),$ with 
\begin{align}
 V_R(t)&=U\dt k\,Z_\mrm c\left[\cos(\m t-\f)\cos(\w_\mrm{rf} t+\dt)+\frac{\w_\mrm{rf}}{\w_z}\cos(\m t-\f)\sin(\w_z t)\sin\dt
  -\cos(\m t-\f)\cos(\w_z t)\cos \dt\right] \sum_n\s^z_n,	
  \label{eqnVR}
\end{align}
and $Z_\mrm c\equiv\ve/[M(\w^2_\mrm{rf}-\w^2_z)]$. Because of the choice of $\al_1(t)$ made in Eq.~\eq{eqnalpha1}, $H_R(t)$ does not explicitly depend on $H_\mrm{rf}(t)$.

We further simplify $H_R(t)$ by performing some assumptions and approximations. Specifically, we take $\mu$ to be tuned to $\mu=\w_\mrm{rf}$ and, assuming sufficiently stable phases, let $\phi=-\dt=0$. In addition, we assume that $\mu=\w_\mrm{rf}$ is much closer to $\w_z$ than to any other mode frequency, and thus neglect all modes other than the COM mode in $H_R(t)$. This gives
\begin{align}
 H_R(t)&\simeq\hbar\w_z a^\dag a+U\dt k\sqrt{\frac\hbar{2MN\w_z}}\cos(\m t)(a+a^\dag)\sum_n\s^z_n -U\sin(\m t)\sum_n\s^z_n\notag\\
  &\hspace*{-7mm}\qquad+U\dt k Z_\mrm c[1+\cos(2\m t)-\cos(\Sg t)-\cos(D t)]\sum_n \frac{\s^z_n}2,	\label{eqnHRsimplified}
\end{align}
where we have dropped the COM mode index by taking $a\equiv a_1$, and introduced $D\equiv \w_z-\mu$ and $\Sg\equiv\w_z+\mu$. Finally, in Eq.~\eq{eqnHRsimplified} we invoke rotating-wave approximations to neglect terms oscillating at frequencies $\mu\gg U/\hbar$, $2\mu\gg U\dt k Z_\mrm c/\hbar$, and $\Sg\gg U\dt k Z_\mrm c/\hbar$, and then find
\begin{align}
 H_R(t)\simeq \frac12 \bigg[U\dt k Z_\mrm c[1-\cos(Dt)]+2U\dt k\sqrt{\frac\hbar{2MN\w_z}}\cos(\m t)(a+a^\dag)\bigg]\sum_n\s^z_n+\hbar\w_z a^\dag a.
\end{align}
Moving to the interaction picture with respect to $H_\mrm B=\hbar \w_z a^\dag a$ and taking $b\equiv U\dt k Z_\mrm c/\hbar$, $y_0(t)=1-\cos(Dt)$, $y(t)=\cos(\mu t)$, and $B(t)=2g[a^\dag\exp(i\w_z t)+\mrm{H.c.}]$, with $\hbar g=U\dt k\sqrt{\hbar/2MN\w_z}$, then completes the mapping to Eq.~(1) of the main text. Assuming that the bath is initially in thermal equilibrium, $B(t)$ then describes a stationary and Gaussian quantum noise process with zero mean, enabling us to use the approach presented in the main text to evaluate $\D b(t)$.

Note that the rotating-wave approximations made above are well justified for typical parameter values of interest. 
From Ref.~\cite{gilmore2017amplitude}, we typically have $U/\hbar\lesssim2\pi\times10$~kHz $\ll \mu\simeq\w_z\simeq2\pi\times1.57$~MHz. In addition, taking $Z_\mrm c\leq10$~nm and $U\dt k\lesssim 40$~yN 
results in $U\dt k Z_\mrm c/\hbar\lesssim2\pi\times600$~Hz $\ll\Sg\simeq2\mu\simeq2\pi\times3.14$~MHz.
Importantly, in order for our mapping to Eq.~(1) to be accurate, it is also essential that non-collective noise mechanisms 
be negligible, or else different irreducible representations with $J <N/2$ would be populated and mixed in the course of the dynamics, even for the symmetric initial states of choice. 
Uncorrelated spontaneous emission, in particular, will ultimately cause leakage outside the initially populated $J=N/2$ subspace. 
Based on the parameters below Eq. (25) of the Supplementary Material of Ref.~\cite{gilmore2017amplitude}, such spontaneous emission occurs with a decay time $\simeq 13$ ms. The dominant correlated quantum noise source we consider, however, acts on a much shorter timescale: more quantitatively, the optimal detection times corresponding to the uncertainties presented in Fig.~2 of our work are all well below 1 ms.  This legitimates neglecting the effects of spontaneous emission.


\bibliographystyle{apsrev4-1}
\bibliography{paper}